\documentclass[letterpaper,titlepage,11pt]{article}
\usepackage{amssymb,amsmath,amsfonts}
\usepackage{epsfig}
\usepackage{graphicx}

\newcommand{\Z}{\mathbb{Z}}

\newcommand{\spa}{\quad , \quad }

\newcommand{\al}{\alpha} 
 \newcommand{\be}{\beta}

 \newcommand{\ta}{\tau}
\newcommand{\om}{\omega} 
 \newcommand{\De}{\Delta}
\newcommand{\La}{\Lambda} \newcommand{\tha}{\theta}

\newcommand{\eq}{\begin{equation}}
\newcommand{\eqx}{\end{equation}}
\newcommand{\eqn}{\begin{eqnarray}}
\newcommand{\eqnx}{\end{eqnarray}}
\newcommand{\f}[2]{\frac{#1}{#2}}

\def\XXint#1#2#3{{\setbox0=\hbox{$#1{#2#3}{\int}$}
     \vcenter{\hbox{$#2#3$}}\kern-.5\wd0}}

\DeclareMathOperator{\tr}{Tr}

\newcommand{\scalea}{A}
\newcommand{\scaleb}{B}
\newcommand{\Lanew}{\La_{\rm new}}

\newcommand{\nn}{{\cal N}}

\newcommand{\Dl}{\Delta}

\newcommand{\Lm}{\Lambda}
\newcommand{\eps}{\varepsilon}

\numberwithin{equation}{section}

\begin{document}

\begin{titlepage}

\rightline{\vbox{\small\hbox{\tt hep-th/0701087} }} \vskip 2.5cm

\centerline{\Large \bf Factorization of Seiberg-Witten Curves }
\vskip .3cm
\centerline{\Large \bf with Fundamental Matter }

\vskip 1.6cm \centerline{\bf Romuald A. Janik$^a$, Niels A.
Obers$^b$ and Peter B. R\o nne$^b$} \vskip 0.5cm

\centerline{\sl  $^a$ Institute of Physics, Jagellonian University}
\centerline{\sl Reymonta 4, 30-059, Krakow, Poland}
\vskip 0.5cm
\centerline{\sl$^b$The Niels Bohr Institute} \centerline{\sl Blegdamsvej 17, 2100
Copenhagen \O, Denmark}

\vskip 0.5cm

\centerline{\small\tt ufrjanik@if.uj.edu.pl, obers@nbi.dk,  roenne@nbi.dk}

\vskip 1.6cm

\centerline{\bf Abstract} \vskip 0.2cm \noindent
We present an explicit construction of the factorization of
Seiberg-Witten curves for $\nn=2$ theory with fundamental flavors.
We first rederive the exact results for the case of complete factorization,
and subsequently derive new results for the case with breaking of gauge symmetry
$U(N_c) \to U(N_1) \times U(N_2)$. We also show that integrality of periods
is necessary and sufficient for factorization in the case of general gauge symmetry
breaking. Finally, we briefly comment on the relevance of these results for the
structure of $\nn=1$ vacua.

\end{titlepage}


\section{Introduction}

The study of supersymmetric gauge theories has provided many new
insights into non-perturbative phenomena in gauge theories. The
constraints imposed by supersymmetry combined with other
simplifications and symmetries have made it possible to obtain exact
non-perturbative results for at least some quantities in these gauge
theories. In particular, in the seminal work of Seiberg and
Witten \cite{Seiberg:1994rs,Seiberg:1994aj} it was shown that the
low-energy dynamics of ${\cal{N}}=2$ supersymmetric gauge theories
is encoded in the properties of an associated hyperelliptic
Seiberg-Witten (SW) curve.

Furthermore, remarkable progress has been made in understanding
the structure of ${\cal{N}}=1$ theories obtained from breaking the
${\cal{N}}=2$ theory by  turning on a superpotential for the adjoint
${\cal{N}}=1$ chiral superfield. Motivated by constructions in
string theory \cite{Cachazo:2001jy}, Dijkgraaf and Vafa
\cite{Dijkgraaf:2002dh} found a link between the effective
superpotentials in these theories and random matrix theory, which
was later understood in purely field theoretic terms
\cite{Ferrari:2002jp,Dijkgraaf:2002xd,Cachazo:2002ry}. (See e.g.
Refs.~\cite{Argurio:2003ym,Ronne:2004qy,Marino:2004eq,Kennaway:2004kw}
for pedagogical introductions)

Much of the physics of the ${\cal{N}}=2$ theory deformed by a
tree-level superpotential can be obtained effectively from the
knowledge that the SW-curve of the undeformed theory factorizes,
since the resulting ${\cal{N}}=1$ theories occur in the region of
the ${\cal{N}}=2$ moduli space where (some) monopoles become
massless. The relation of this to the matrix model conjecture of
Dijkgraaf and Vafa were studied in e.g.
Refs.~\cite{Cachazo:2002pr,Ferrari:2002jp,Gopakumar:2002wx,Naculich:2002hi,%
Janik:2002nz,Balasubramanian:2002tm,Boels:2003at}.

In these developments one has
primarily focussed on the case in which the gauge group is not broken, using
factorization of the SW-curve in terms of Chebyshev polynomials as found
in \cite{Douglas:1995nw}. If one goes beyond this, and considers for example
the breaking $U(N_c) \rightarrow U(N_1) \times U(N_2)$ it was found
\cite{Cachazo:2002zk,Ferrari:2002kq,Ferrari:2003yr} that
the space of ${\cal{N}}=1$ vacua exhibits a very complex structure of various
connected components, each of which allows for multiple dual descriptions of the
same physics but with different patterns of breaking. While in these works
one had to consider the factorization of SW-curves on a case by case basis
for low $N_c$, Ref.~\cite{Janik:2003hk} contains an exact solution of
the factorization problem for arbitrary $N_c$ for any gauge breaking
of the form $U(N_c) \rightarrow U(N_1) \times U(N_2)$.
This was then also used  to further study the
global structure of ${\cal{N}}=1$ vacua (see also \cite{Janik:2005sk}).

Although the case without flavors already exhibits a very rich structure,
inclusion of flavors in the circle of ideas discussed is of physical interest
and has received a lot of attention as well. In particular, matrix models methods
were used in Ref.~\cite{Demasure:2002jb} to obtain a solution of the
complete factorization of SW-curves for theories with fundamental
flavors. See also
Refs.~\cite{Argurio:2002xv,Seiberg:2002jq,Naculich:2002hr,Klein:2003we,%
Ahn:2004ym,Gomez-Reino:2004dr,Elmetti:2005sh} for work on the
relation to the matrix model in the presence of fundamental matter.
Following the work \cite{Cachazo:2002zk} the phase structure of
${\cal{N}}=1$ theories with fundamental matter was then explored in
Refs.~\cite{Cachazo:2003yc,Balasubramanian:2003tv,Ahn:2003ui,Ahn:2003vh,
Merlatti:2003iy}. Also here, one is confined to considering specific
cases with low $N_c$ and low number of flavors $N_f$

The purpose of this paper is to extend the factorization of the
SW-curve for the gauge breaking $U(N_c) \rightarrow U(N_1) \times
U(N_2)$ found in \cite{Janik:2003hk} to the case when fundamental
matter is included in the ${\cal{N}}=2$ supersymmetric $U(N_c)$ gauge
theory. An exact expression of the factorization of the SW-curve in
this genus one case is obtained for arbitrary $N_c$ and $0 \leq N_f
< 2 N_c$. The construction includes the genus one case of
\cite{Janik:2003hk} for $N_f=0$. Furthermore, it correctly reduces
to the genus zero case with flavors in the fundamental, reproducing
the results of Ref.~\cite{Demasure:2002jb} in a simpler way. As an
important ingredient we prove that integrality of periods is
necessary and sufficient for factorization in the case of general
gauge symmetry breaking.

The structure of the paper is as follows. In Section 2 we briefly review
some useful facts about
${\cal{N}}=2$ supersymmetric QCD for $U(N_c)$ gauge theory with $N_f$ flavors
transforming in the fundamental, including the corresponding SW-curve.
Section 3 presents the factorization of the SW-curve, and defines in particular
the problem of the factorization when $N_c-2$ monopoles in the low
energy effective action become massless. This corresponds to the gauge group
breaking $U(N_c) \rightarrow U(N_1) \times U(N_2)$ and the physics will
only depend on a reduced curve which is elliptic, i.e has genus one.
In Section 4 we summarize the equations that the meromorphic one-form
needs to obey in order to solve the factorization problem. We then consider
in Section 5 first the genus zero case of complete factorization,
where we rederive in a very simple way the original results of \cite{Demasure:2002jb}.

Section 6 then contains the main result of the paper, namely the exact solution
of the factorization of the SW-curve in the genus one case when fundamental
matter is present.
We end with the conclusions and open problems in Section 7. Two
appendices are included. In Appendix A we prove the statement that a
necessary and sufficient condition for factorization of the SW-curve
is integrality of the periods as specified in Section 3. Appendix B
considers the general solution of Section 6 when flavors are decoupled,
reproducing the genus one solution of \cite{Janik:2003hk}.


\section{$\mathcal{N}=2$ SQCD}

We consider an $\mathcal{N}=2$ supersymmetric $U(N_c)$
gauge theory with $N_f$ flavors, i.e. we add the following terms to
the pure $\mathcal{N}=2$ SYM Lagrangian for the (adjoint) chiral
superfield, $\Phi$ \cite{Alvarez-Gaume:1996mv}
\begin{equation}\label{eqn1}
    \int\!\mathrm{d}\tha^4\left(Q^{\dagger}_ie^{-2V}Q_i+\widetilde{Q}_ie^{-2V}\widetilde{Q}^{\dagger}_i\right)+\int\!\mathrm{d}\tha^2\left(\sqrt{2}\widetilde{Q}_i\Phi Q_i+\sqrt{2}\widetilde{Q}_iM_{ij}Q_j\right) +
    \mathrm{h.c.} \ .
\end{equation}
Here we have suppressed the gauge indices, and $Q_i$
($\widetilde{Q}_i$) are chiral superfields transforming in the
(anti-)fundamental representation of $U(N_c)$. The flavor
index $i$ runs from 1 to $N_f$. The mass matrix fulfills
$[M,M^{\dagger}]=0$ so it can be diagonalized \cite{Argyres:1995wt}
by a rotation in flavor-space. Denoting the eigenvalues of the mass
matrix as $m_i$, the superpotential in \eqref{eqn1} takes the form
\begin{equation}\label{eqn2}
\sqrt{2}\widetilde{Q}_i\Phi Q_i+\sqrt{2}m_i\widetilde{Q}_iQ_i.
\end{equation}

The object of our interest is the vacuum moduli space for the
Coulomb phase of the theory where, classically, we have zero
expectation values for the quarks, $\Phi$ is diagonal, and its
eigenvalues $\phi_a$ parameterize the moduli space. Generically,
$U(N_c)$ breaks down to $U(1)^{N_c}$ but if some
of the eigenvalues $\phi_a$ coincide we get non-abelian factors.
Further, if $\phi_a+m_i=0$ we get a massless quark. Quantum
mechanically, the vacuum moduli space is $N_c$-dimensional and
parameterized by
\begin{equation}\label{eqn3}
    u_k=\frac{1}{k}\left<\tr\Phi^k\right>,\quad k=1,\ldots,N_c \ .
\end{equation}
These are encoded in the polynomial
\begin{equation}\label{eqn4}
P_{N_c}(x,u_k)=\left<\mathrm{det}(xI-\Phi)\right>=x^{N_c}+\sum_{i=1}^{N_c}s_i
x^{N_c-i} \ ,
\end{equation}
where the coefficients, $s_i$, are polynomials in the $u_k$'s and
the relation is given by Newton's formula
\begin{equation}\label{eqn5}
    is_i+\sum_{k=1}^{i}ks_{i-k}u_k=0,\qquad i=0,\ldots,N_c \ ,
\end{equation}
where we define $s_0\equiv1$.

As is well-known, without fundamental matter the low energy
effective description of the theory is beautifully captured by a
one-form on a hyperelliptic curve, the Seiberg-Witten (SW) curve, of
genus $N_c-1$ \cite{Seiberg:1994rs,Seiberg:1994aj}. In the case with
fundamental matter, the SW-curve takes the form
\begin{equation}\label{eqn6}
    y^2=P_{N_c}(x,u_k)^2-4\La^{2N_c-N_f}\prod_{i=1}^{N_f}(x+m_i),\qquad
    N_f<2N_c \ ,
\end{equation}
where $m_i$ are the bare masses from~\eqref{eqn2}. This was first
found for the $\mathrm{SU}(2)$ gauge group
(see \cite{Seiberg:1994aj} and references therein) and later for
general $\mathrm{SU}(N)$ gauge groups
in \cite{Argyres:1995wt,Hanany:1995na}.  For $N_f\geq N_c$ the
curve is, however, ambiguous and we can add a polynomial to
$P_{N_c}$ without changing the prepotential of the low-energy
effective theory (see Refs.~\cite{Hanany:1995na,D'Hoker:1996nv})
(for $N_c\leq N_f<2N_c$)
\begin{equation}\label{eqn7}
    y^2=(P_{N_c}(x,u_k)+\La^{2N_c-N_f}Q_{N_f-N_c}(x,m_i,\La))^2-4\La^{2N_c-N_f}\prod_{i=1}^{N_f}(x+m_i)
     \ ,
\end{equation}
where $Q_{N_f-N_c}$ is a polynomial of degree $N_f-N_c$
independent of the $u_k$'s.

In this paper we will use the SW-curve in the
form~\eqref{eqn6}. Note that we can lower the value of
$N_f$ with one unit by taking the limit $m_i\rightarrow\infty,\,\La\rightarrow0$
for a given $i$ while keeping
$\La^{2N_c-N_f}m_i\equiv\Lanew^{2N_c-(N_f-1)}$ constant. This
corresponds to removing the $i$th flavor and the new scale $\Lanew$
is obtained by scale matching. In this way one can find
the remaining curves for lower $N_f$, given the curve for $N_f=2N_c$, by taking the appropriate
limits \cite{Argyres:1995wt}. In particular, using this procedure
one can also obtain the curve without fundamental matter.
The constraint $N_f<2N_c$ is needed for the theory to be
asymptotically free (see e.g. \cite{D'Hoker:1999ft}) and the metric
on our moduli space will otherwise not be positive definite for
large $\Phi$ \cite{Cachazo:2003yc}.

\section{Factorization}

We will investigate submanifolds of the moduli space where the
SW-curve factorizes, i.e. it takes the form
\begin{equation}\label{eqn8}
    y^2=P_{N_c}(x,u^{({\rm fact})}_k)^2-4\La^{2N_c-N_f}\prod_{i=1}^{N_f}(x+m_i)=
    F_{2(N_c-n)}(x)H_{n}(x)^2\ ,
\end{equation}
where $F_{2(N_c-n)}$ and $H_{n}$ are polynomials of degree
$2(N_c-1)$ and $n$, respectively. Here $n$ denotes the number of
double roots for the curve. We will assume there are no multiple
roots in $F_{2(N_c-n)}$ and $H_n$ and that they have no common roots, i.e.
we have no roots of order higher than two. As in the case without fundamental matter,
we expect that the solution for a given $n$ should constrain $n$ of the $N_c$ free parameters so
we end up with $N_c-n$ continuous parameters. However, the subspace of
factorized solutions is not invariant under translations of $x$ as
in the case without matter, since such a translation would simply change
the masses $m_i$. We thus  have to look for this continuous parameter
elsewhere.

These points of factorization are of special interest since here $n$
of the ${N_c-1}$ monopoles in the low energy effective action become
massless. Importantly, these are the points we will be localized at
when we softly break the $\mathcal{N}=2$ supersymmetry to
$\mathcal{N}=1$ by addition of a tree-level superpotential for
$\Phi$. Here the gauge groups breaks into $N_c-n$ parts.

In the case without fundamental matter and complete factorization,
i.e. when $n=N_c-1$ and we only have two single roots, the problem
was solved in Ref.~\cite{Douglas:1995nw} using Chebyshev polynomials.
This case corresponds to an unbroken gauge group in the low energy
effective theory. If we have $n=N_c-2$, i.e. four single roots, the
general factorization was obtained using the elliptic theta function
in \cite{Janik:2003hk} (see also \cite{Janik:2005sk}).

\begin{figure}
\begin{center}
{\includegraphics{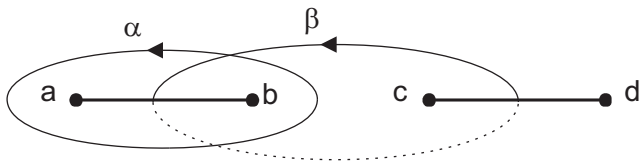}}
\end{center}
\caption{The figure shows the cycles and cuts for the elliptic curve. We assume that
        $F_{4}(x)=(x-a)(x-b)(x-c)(x-d)$. The $\al$ and $\be$
        cycles are shown. A dotted line means that the curve is on
        the lower sheet. Note that $\al$ and $\be$ only
        intersect once.}
\label{fig1}
\end{figure}

Including fundamental matter the complete factorization case was
solved in \cite{Demasure:2002jb} (see also \cite{Janik:2003kf}
and \cite{Kennaway:2004kw}). In this paper we will solve the
$n=N_c-2$ case where we have four single roots following closely the
approach in \cite{Janik:2003hk}. In this case the
factorization~\eqref{eqn8} takes the form
\begin{equation}\label{eqn9}
    y^2=P_{N_c}(x,u^{({\rm fact})}_k)^2-4\La^{2N_c-N_f}\prod_{i=1}^{N_f}(x+m_i)=
    F_{4}(x)H_{N_c-2}(x)^2  \ .
\end{equation}
The solution should be parameterized by 2 continuous parameters and
a set of discrete parameters, as we indeed will see. Importantly,
the physics will only depend on the reduced
curve \cite{Cachazo:2002pr}
\begin{equation}\label{eqn10}
    y_{\rm red}^2=F_{4}(x) \ ,
\end{equation}
which is elliptic, i.e. has genus one, and is represented by a torus.
Due to the square in~\eqref{eqn10} we can, as usual, see the curve
as represented by two sheets connected by two cuts between the four
roots of $F_4$. The sheets are compactified by adding points at
the infinities thus giving us a torus. On this torus we have a
canonical homology basis consisting of the cycles $\al$ and $\be$.
See Figure~\ref{fig1} where the cuts also are shown.

\section{Construction of the solution}

As in Ref.~\cite{Janik:2003hk} the idea is to consider
\begin{equation}\label{eqn11}
    \om\equiv
    T(x)\mathrm{d}x\equiv\left<\tr\frac{\mathrm{d}x}{x-\Phi}\right> \ ,
\end{equation}
which is a meromorphic one-form on the SW-curve. If we
can determine $\om$ we have, in principle, solved the problem
since the $u_k$'s can then be obtained to any order using the relation
\begin{equation}\label{eqn11.1}
u_k=-\frac{1}{k}\mathrm{res}_{x=\infty}\,x^k\om \ ,
\end{equation}
where we have defined
\begin{equation}\label{eqn11.2}
    \mathrm{res}_{x=\infty}\,\frac{1}{x}\mathrm{d}x=-1 \ .
\end{equation}
As was shown in \cite{Cachazo:2003yc} $\om$ has residue $-N_c$ at
infinity of the upper sheet, residue $N_c-N_f$ at the infinity of
the lower sheet, and residues $+1$ at $-m_i$ (the zeroes of
$\prod_i(x+m_i)$) on the lower sheet.\footnote{If some of the masses
coincide the residues should simply be added.} One can just as well
have some of the mass-poles on the upper sheet and still solve the
factorization problem. This will then correspond to Higgs vacua
rather than Coulomb vacua \cite{Cachazo:2003yc}.
In the quantum theory there is no physical boundary between the two
sheets.

Furthermore, $\om$ has integral periods
\begin{equation}\label{11.3}
    \frac{1}{2\pi i}\int_{\al}\om=N_1,\quad\frac{1}{2\pi i}\int_{\be}\om=\Delta k \ ,
\end{equation}
where $N_1$ and $\Delta k$ are integers, and we have to think of
definite curves for $\al$ and $\be$ only encircling the cuts and not
any of the poles of $\om$.  The integrality of the periods is a
necessary and sufficient condition for the factorization of the
SW-curve as will be made precise and proven in
Appendix~\ref{app:theorem}. This is actually independent of the
number of cuts in the factorization. As was also shown
in \cite{Cachazo:2003yc} this means that $\om$ takes the form
\begin{equation}\label{eqn12}
    \om=\mathrm{d}\log(P_{N_c}(x)+y(x))=\bigg(\frac{P'_{N_c}(x)}{y(x)}+
    \frac{B'(x)}{2B(x)}-\frac{P_{N_c}(x)B'(x)}{2y(x)B(x)}\bigg)\mathrm{d}x \ ,
\end{equation}
where we have defined $B(x)=\prod_i(x+m_i)$. Here the residues can
easily be checked and the integral periods follows from $\om$
being the derivative of a logarithm. From~\eqref{eqn12} we see
that not only can we retrieve the $u_k$'s from $\om$ but we can
also get $\La$:
\begin{equation}\label{eqn12.5}
    \int_{\widetilde{\La_0}}^{\La_0}\om=-\log(\La^{2N_c-N_f})+\log(\La_0^{2N_c-N_f})+\mathcal{O}\!\left(\frac{1}{\La_0}\right),
\end{equation}
where we think of $\La_0$ ($\widetilde{\La_0}$ is the corresponding
point on the lower sheet) as a large cut-off for the integration
$\int_{\infty_-}^{\infty_+}\om$. Here $\infty_{\pm}$ refers to the
infinities on the upper/lower sheet.

Before we proceed to the genus one case which is the main focus of this
paper let us exhibit the construction in the simpler case of genus zero
(no gauge symmetry breaking),
where we rederive in a very simple way the factorization formulas
of Ref.~\cite{Demasure:2002jb}.

\section{Genus zero case}

In the genus zero case we have $N_c-n=1$ and we expect a single
continuous parameter in the solution.

Let us start from the reduced curve which in this case is given by
the equation
\eq \label{e.ftwo} y^2=F_2(x) \equiv (x-a)(x-b) \ .
\eqx
As
explained in the previous section, we have to construct a
meromorphic 1-form $\omega$ on the curve with residues $-N_c$ at
infinity on the physical sheet, $N_c-N_f$ at infinity on the second
sheet and with residue 1 at $x=-m_i$.

It turns out to be much easier to use an unconstrained parametrization
of the reduced curve, i.e. to pass to the universal covering space.

\subsubsection*{Parametrization and $\Z_2$ map.}

Since the curve (\ref{e.ftwo}) has genus zero, it can be parameterized
by functions on a sphere, which is represented as a compactified
complex plane. This can be done very easily. Let us first rewrite
the equation (\ref{e.ftwo}) in the form
\eq y^2=(x-T)^2-4R \ ,
\eqx
where we used the notation of \cite{Demasure:2002jb}
\eq
T=\f{a+b}{2} \spa R=\f{(a-b)^2}{16} \ .
 \eqx
 Then a rational parameterization is
\begin{equation}
x  =  T+ 2 \sqrt{R} \f{1+z^2}{1-z^2} \quad, \quad
y  =  2\sqrt{R} \f{2z}{1-z^2}\ .
\end{equation}

For our application we have to keep track of some additional
structure on the curve. Firstly, we have to single out points on the
sphere $\infty_+$, $\infty_-$ which correspond to points at infinity
in the $(x,y)$ plane. Here these are $z=\pm 1$. Secondly, it is
convenient to exhibit the $\Z_2$ covering transformation which
exchanges the sheets $(x,y) \to (x,-y)$. In terms of the $z$
coordinate it is represented as $z \to -z$. Its fixed points are
exactly the branch points of the curve \eqref{e.ftwo}. These are
$z=0$ and $z=\infty$ and correspond to $x=T+2\sqrt{R}$ and
$x=T-2\sqrt{R}$ respectively.

\subsubsection*{The meromorphic 1-form}

Using the $z$ coordinate we can at once write the unique meromorphic
1-form with the prescribed poles and residues
\eq
\omega=\left(\f{-N_c}{z-1}+ \f{N_c-N_f}{z+1}
+\sum_{i=1}^{N_f}\f{1}{z+z_{m_i}} \right) dz \ ,
\eqx
where the
location of the pole corresponding to $x=-m$ can be found to be \eq
z_m=\pm \f{\sqrt{(m+T)^2-4R}}{m+T-2\sqrt{R}} \ . \eqx The two choices
of sign correspond to putting the pole on either of the two sheets.
Since all parameters are complex we can always analytically continue
the answer from one sheet to the other one. As mentioned
above, this has the interpretation of interpolating between Coulomb and Higgs
vacua.

\subsubsection*{Factorization solution}

We can now calculate the $u_k$'s using (\ref{eqn11.1}). Remarkably
enough all the formulas from \cite{Demasure:2002jb} (compare e.g.
formulas (38)-(40) in \cite{Janik:2003kf}) now follow from the
simple formula \eq u_k=-\frac{1}{k}\mathrm{res}_{x=\infty}\,x^k\om
=-\f{1}{k} \mathrm{res}_{z=1} \left( T+ 2 \sqrt{R} \f{1+z^2}{1-z^2}
\right)^k \cdot \om \ . \eqx
The final ingredient is the calculation of
$\Lambda$. We use formula (\ref{eqn12.5}) in the form: \eq \log
\Lm^{2N_c-N_f}= -\lim_{\eps \to 0} \left\{ \int_{-1+\eps}^{1-\eps}
\om - N_c \log x(1-\eps)- (N_c-N_f) \log x(-1+\eps) \right\} \ . \eqx
After a brief calculation one gets
\eqn\label{eqlambdagenuszero}
\Lm^{2N_c-N_f} &=& R^{N_c-\f{N_f}{2}} \bigg(\f{\sqrt{(m+T)^2-4R}-m-T
  + 2\sqrt{R}}{\sqrt{(m+T)^2-4R}+m+T- 2\sqrt{R}}\bigg)^{N_f} \nonumber\\
&=& \f{R^{N_c}}{\prod_{i=1}^{N_f} \f{1}{2} \left( m+T+
  \sqrt{(m+T)^2-4R} \right)} \ ,
\eqnx which is exactly the formula obtained from matrix models
in \cite{Demasure:2002jb}. Plugging these parameters into the
SW-curve will lead to a complete factorization
regardless of whether the flavor poles are on a single or on
different sheets (which amounts to a choice of the signs of the
relevant square-roots). Equation~\eqref{eqlambdagenuszero} exactly
gives one constraint so we end up with one continuous parameter as
expected.

\subsubsection*{Number of vacua}

Finally, let us discuss the number of such vacua. From the above
construction one can obtain a discrete set of $2N_c-N_f$ vacua in
the following manner. Let us rescale the parameters by
\eq T \to
e^{i\al} T \spa R \to e^{2i\al} R \spa  m \to e^{i\al} m \ .
\eqx
Then $x$ is effectively rescaled as $x \to e^{i\al} x$. In order for
the resulting factorization to be related to the same theory,
$\Lm^{2N_c-N_f}$ should be unchanged hence \eq \al=2\pi
\f{k}{2N_c-N_f}\spa k = 0 , \ldots, 2 N_c - N_F-1 \ , \eqx
which proves the claim.

\section{Genus one case}

Let us now adopt the same strategy in our main case of interest
i.e. the genus one case. This case is especially interesting as, in
contrast to the genus zero described above, there is gauge symmetry
breaking, one has additional discrete parameters labelling the vacua
(inequivalent factorizations), new types of Coulomb vacua appear with
increasing $N_c$ which cannot be induced from those with smaller
$N_c$. Even more interestingly, differing discrete labels like
$(N_1,N_2,\Dl k,k) \neq (N'_1,N'_2,\Dl k',k')$ may lead to {\em the
  same} factorized SW-curves thus allowing for {\em dual}
descriptions of the same physics.

Here we start from the reduced curve which in this case is given by
the quartic equation \eq \label{e.ffour} y^2=F_4(x) \equiv
(x-a)(x-b)(x-c)(x-d)\ . \eqx Note that for cubic superpotentials $W(x)$ relevant
to this case the right hand side can be written as $F_4(x)=W'(x)^2-f(x)$ with $f(x)$
a linear polynomial.

\subsubsection*{Parametrization and $\Z_2$ map.}

Since the above curve is quartic, it has genus one and hence can be
parameterized by a torus i.e. the complex plane modulo $(1,\tau)$,
where $\ta$ is a complex parameter (the modulus) with positive
imaginary part.

Again we would like to exhibit the $\Z_2$ covering map (the
hyperelliptic involution) moving between the two branches of
(\ref{e.ffour}). A convenient choice is
\eq z \to 1+\tau-z \ .
\eqx
There are {\em four} fixed points on the torus: $0$, $1/2$, $\tau/2$
and $1/2+\tau/2$ which under the embedding $x(z)$, which we will soon
give explicitly, go over to the branch points of
(\ref{e.ffour}).

\begin{figure}
\begin{center}
{\includegraphics{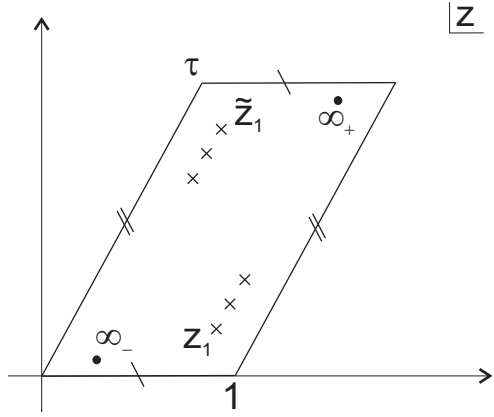}}
\end{center}
\caption{The figure shows the torus with  modular parameter
        $\tau$ and the two points $\infty_+$ and $\infty_-$
        corresponding to the infinities, denoted by dots, on the upper and lower
        sheet. Also shown are the points $z_i$, denoted by crosses, corresponding to
        the masses $m_i$. $\tilde{z}_i$ are the corresponding
        points on the lower sheet.}
\label{fig2}
\end{figure}

Next we
need to mark the two points corresponding to the infinities on the
upper and lower sheet -- these will be denoted $\infty_+$ and
$\infty_-$ respectively. In Figure~\ref{fig2} the torus is
illustrated with the two marked points.

We should require that the points at infinity go to each other under
the $\Z_2$ covering map thus giving the relation
\eq
\label{infrel}
\infty_+=1+\tau-\infty_- \ . \eqx
We will find that then $\infty_-$
will be fixed completely when constructing the meromorphic 1-form.

\subsubsection*{The meromorphic 1-form}

Let us now construct the meromorphic 1-form with the appropriate
properties. The form has to have poles at $z=\infty_\pm$ and $z_i$
such that $x(z_i)=-m_i$ (see Figure~\ref{fig2}) with the prescribed residues
and integral periods:
\begin{gather}
    \mathrm{res}_{z=\infty_+}\,\om=-N_c,\quad\mathrm{res}_{z=\infty_-}\,
     \om=N_c-N_f,\quad\mathrm{res}_{z=\tilde{z}_i}\,\om=1,\label{eqn13}\\
    \frac{1}{2\pi i}\int_{0}^1\om=N_1,\quad\frac{1}{2\pi
    i}\int_{0}^{\tau}\om=\Delta k \ ,\label{eqn14}
\end{gather}
where we have used that the $\al$-cycle integration corresponds to
integrating from $0$ to $1$ on the torus, and the $\be$-cycle from
$0$ to $\ta$. We again have to think of definite curves for the
integration not encircling the poles of $\om$.

On the torus we have unique one-forms $\om_{PQ}$ which have zero
$\alpha$ period and simple poles in $P$ and $Q$ with residues $+1$
and $-1$ respectively. Using these we can write $\om$ uniquely:
\begin{equation}\label{eqn16}
    \om=N_c\om_{\infty_-\infty_+}+\sum_i\om_{\tilde{z}_i
    \infty_-}+2\pi i N_1\mathrm{d}z \ ,
\end{equation}
where we have used that after determining the residues the only
redundancy is in the addition of a holomorphic one-form which on the
torus has the simple form of a constant times $\mathrm{d}z$. The
constant is then determined by the $\al$-period in~\eqref{eqn14}.
The remaining $\be$-period gives a constraint determining $\infty_+$
(which is equal to $-\infty_-$ modulo $(1,\ta)$). Thus it seems like
no continuous parameter is undetermined on the torus except the
modular parameter $\tau$. However, for the embedding we have a
scalar factor and a translation. We will see that we determine the
scale factor using~\eqref{eqn12.5} thus leaving us with two
continuous parameters.

Now, the main point is that on the torus we have a formula for
$\om_{PQ}$ using the elliptic theta function:
\begin{equation}\label{eqn17}
    \tha(z,\tau)=\sum_{n=-\infty}^{\infty}e^{i\pi\tau n^2+i2\pi
    zn} \ .
\end{equation}
We will often suppress the $\ta$ dependence and just write
$\tha(z)$. The theta function is a multiplicative holomorphic
function with periods
\begin{equation}\label{eqn18}
    \tha(z+1,\tau)=\tha(z,\tau),\quad \tha(z+\tau,\tau)=e^{-i\pi\ta-i2\pi
    z}\tha(z,\tau) \ .
\end{equation}
Importantly, $\tha(z)$ is only zero in $(1+\ta)/2$ and the
multiplicity is one. Using~\eqref{eqn18} this gives us the formula
for $\om_{PQ}$ (see e.g. \cite{Mumford:1983})
\begin{equation}\label{eqn19}
    \om_{PQ}=\mathrm{d}\log\frac{\tha(z-P+\tfrac{1+\ta}{2})}{
    \tha(z-Q+\tfrac{1+\ta}{2})} =\frac{\tha'(z-P+\tfrac{1+\ta}{2})}{
    \tha(z-P+\tfrac{1+\ta}{2})} \mathrm{d}z-\frac{\tha'(z-Q+
    \tfrac{1+\ta}{2})}{\tha(z-Q+\tfrac{1+\ta}{2}) }\mathrm{d}z  \ .
\end{equation}
Using~\eqref{eqn16} we thus get an explicit expression for $\om$:
\begin{equation}\label{eqn20}
    \om=N_c\mathrm{d}\log\frac{\tha(z-\infty_-+\tfrac{1+\ta}{2})}{
    \tha(z-\infty_++\tfrac{1+\ta}{2})}+ \sum_i\mathrm{d}\log\frac{
    \tha(z-\tilde{z}_i+ \tfrac{1+\ta}{2})}{\tha(z-\infty_-+
    \tfrac{1+\ta}{2})}+2\pi i N_1\mathrm{d}z \ .
\end{equation}
We should now perform the $\be$-cycle integration
in~\eqref{eqn14}. Using~\eqref{eqn18} this gives
\begin{equation}\label{eqn21}
    -N_c \infty_++(N_c-N_f)\infty_-+\sum_i\tilde{z}_i+N_1\tau=\Delta
    k \ ,
\end{equation}
which also directly follows from $\int_{\be}\om_{PQ}=2\pi i (P-Q)$.
We may now use the relation \eqref{infrel} between $\infty_+$  and $\infty_-$ derived
earlier to obtain
\begin{equation}\label{eqn22}
    \infty_-=\frac{(N_1-N_c)\ta-\De k-N_c+\sum_i\tilde{z}_i}{N_f-2N_c} \ ,
\end{equation}
where we think modulo $(1,\tau)$ on the torus. Note that, as is
suggested by this equation, we could trade in $N_1$ and $\Delta k$
for the location of the flavor poles in appropriate copies of the
fundamental domain.

At this stage we have uniquely fixed the meromorphic 1-form $\om$ and
hence we may now extract the factorization solution.

\subsubsection*{Factorization solution}

The $u_k$'s are given by calculating the residues of $x(z)^k \om$ at
$z=\infty_+$  using~\eqref{eqn11.1}:
\begin{equation}\label{eqn28}
    u_k^{({\rm fact})}=-\frac{1}{k}\mathrm{res}_{z=\infty_+}\,x^k\om \ .
\end{equation}
We thus have to construct the embedding map $x(z)$. It has to be a
meromorphic map with single poles at $\infty_+$, $\infty_-$. Then
necessarily it will have two zeroes $z_0$ and, since it should be
invariant under the $\Z_2$ map, $\tilde{z}_0=1+\tau-z_0$. It is thus
fixed uniquely up to a multiplicative constant $B$ and the embedding map
takes the form
\begin{equation}\label{e.xz}
    x(z)=\scaleb\frac{\tha(z-z_0+\tfrac{1+\ta}{2})\tha(z-1-\ta+z_0+
     \tfrac{1+\ta}{2})}{\tha(z-\infty_++\tfrac{1+\ta}{2})
     \tha(z-\infty_-+\tfrac{1+\ta}{2})} \ ,
\end{equation}

In order to compute the complete solutions it remains to determine
$\Lm$. To this end let us perform the integral in~\eqref{eqn12.5}:
We take $z_{\La_0}$ as the point corresponding to $\La_0$ on the
upper sheet, i.e. we think of $z_{\La_0}$ as being close to
$\infty_+$. Then $\widetilde{z_{\La_0}}=1+\ta-z_{\La_0}$.
Using~\eqref{eqn20} we get
\begin{multline}\label{eqn23}
    \int_{1+\ta-z_{\La_0}}^{z_{\La_0}}\om=-\log
    (z_{\La_0}-\infty_+)^{2N_c-N_f}+\log\tha(\infty_+-\infty_-+
    \tfrac{1+\ta}{2})^{2N_c-N_f}\\
    +\sum_i\log\frac{\tha(\infty_+-\tilde{z}_i+\tfrac{1+\ta}{2})}{
    \tha(\infty_--\tilde{z}_i+ \tfrac{1+\ta}{2})}+\log\tha'(\tfrac{1+
    \ta}{2})^{N_f-2N_c} \\
    +(N_c+N_1)2\pi i(\infty_+-\infty_-)-N_f\pi i+
    \mathcal{O}(z_{\La_0}- \infty_+) \ ,
\end{multline}
where we have used $\tha(-z+\tfrac{1+\ta}{2})=\exp(i\pi+i2\pi
z)\tha(z+\tfrac{1+\ta}{2})$ which can be proven
using~\eqref{eqn18} and that $\tha(z)$ is an even function. Since
$x(z)$ has a pole of order one at $\infty_+$ we can write
\begin{equation}\label{eqn23.5}
    \La_0=x(z_{\La_0})=\scalea\frac{1}{z_{\La_0}-\infty_+}+
    \mathcal{O}((z_{\La_0}-\infty_+)^0) \ ,
\end{equation}
where $\scalea$ is a constant. Thus
\begin{equation}\label{23.6}
    \La_0(z_{\La_0}-\infty_+)=\scalea+\mathcal{O}(z_{\La_0}-\infty_+) \ .
\end{equation}
Hence we get the relation
\begin{equation}\label{eqn23.7}
    \log\La_0^{2N_c-N_f}+\log(z_{\La_0}-\infty_+)^{2N_c-N_f}=\log
    \scalea^{2N_c-N_f}+ \mathcal{O}(z_{\La_0}-\infty_+) \ .
\end{equation}
Using this to equate~\eqref{eqn12.5} and~\eqref{eqn23}, we finally
see that~\eqref{eqn12.5} determines the scale $\scalea$ of $x(z)$:
\begin{multline}\label{eqn23.8}
    \log\scalea^{2N_c-N_f}=\log\La^{2N_c-N_f}+\log\tha(\infty_+-\infty_-+
    \tfrac{1+\ta}{2})^{2N_c-N_f}\\
    +\sum_i\log\frac{\tha(\infty_+-\tilde{z}_i+\tfrac{1+\ta}{2})}{\tha(
    \infty_--\tilde{z}_i+\tfrac{1+\ta}{2})}+
    \log\tha'(\tfrac{1+\ta}{2})^{ -2N_c+N_f}\\
    +(N_c+N_1)2\pi i(\infty_+-\infty_-)-N_f\pi i \ .
\end{multline}
This is solved as
\begin{multline}\label{eqn23.9}
    \scalea=\La e^{\frac{i2\pi
    k}{2N_c-N_f}}\frac{\tha(\infty_+-\infty_-+\tfrac{1+\ta}{2})}{\tha'(
    \tfrac{1+\ta}{2})}\Big(\prod_i\frac{\tha(\infty_+-\tilde{z}_i+
    \tfrac{1+\ta}{2})}{\tha(\infty_--\tilde{z}_i+\tfrac{1+\ta}{2})}\Big)^{
    \frac{1}{2N_c-N_f}}\\
    \times e^{2\pi
    i(\infty_+-\infty_-)\frac{N_c+N_1}{2N_c-N_f}}e^{-\pi
    i\frac{N_f}{2N_c-N_f}} \ ,
\end{multline}
where $k$ is an integer, $k=0,\ldots,2N_c-N_f-1$, which is a
discrete parameter of our solution.

Let us now relate $\scalea$ to the scalar factor $\scaleb$ appearing
in the expression (\ref{e.xz}) for the embedding $x(z)$ using
$\lim_{z\rightarrow\infty_+}x(z)(z-\infty_+)=\scalea$. The resulting expression
for $\scaleb$ is
\begin{multline}\label{eqn27}
    \scaleb=\La e^{\frac{i2\pi
    k}{2N_c-N_f}}\frac{\tha(\infty_+-\infty_-+\tfrac{1+\ta}{2})^2}{
    \tha(\infty_+-z_0+\tfrac{1+\ta}{2})\tha(\infty_+-1-\ta+z_0+
    \tfrac{1+\ta}{2})}\\
    \times
    \Big(\prod_i\frac{\tha(\infty_+-\tilde{z}_i+\tfrac{1+\ta}{2})}{
    \tha(\infty_--\tilde{z}_i+\tfrac{1+\ta}{2})}\Big)^{\frac{1}{2N_c-N_f}}
    e^{2\pi i(\infty_+-\infty_-)\frac{N_c+N_1}{2N_c-N_f}}e^{-\pi
    i\frac{N_f}{2N_c-N_f}} \ .
\end{multline}

Thus we have solved the problem and the solution is summarized by Eqs.~\eqref{eqn20},
\eqref{eqn28}, \eqref{e.xz} and \eqref{eqn27}.
As expected, the construction depends on the two continuous parameters $\ta$ and $z_0$ (modulo
$(1,\ta)$) and the discrete parameters $N_1,\Delta k$ and $k$. The
physical given parameters are $N_c, N_f, \La$ and the masses $m_i$.
In principle we should determine the $\tilde{z}_i$s,
$i=1,\ldots,N_f$, using
\begin{equation}\label{eqn29}
    x(\tilde{z}_i)=-m_i,\quad i=1,\ldots,N_f \ .
\end{equation}
However, the dependence on $\tilde{z}_i$ is extremely complicated
since also $x(z)$ in \eqref{e.xz} depends on the $\tilde{z}_i$s through $\scaleb$
(see \eqref{eqn27}).
There is, however, one exception: If all the masses are the same,
$m_i=m$, and correspondingly $\tilde{z}_i=\tilde{z}_1$. Then we can
consider $x'(z)=x(z)+m$. This is zero in $\tilde{z}_1$ and has the
same poles as $x$. Thus $x'$ is given by~\eqref{e.xz}
and~\eqref{eqn27} with $\tilde{z}_1=z_0$, and~\eqref{eqn28} is
replaced by
\begin{equation}\label{eqn30}
    u_k^{({\rm fact})}=-\frac{1}{k}\mathrm{res}_{z=\infty_+}\,(x'-m)^k\om \ .
\end{equation}
Of course, in the case of different masses we can in the same way
trade $z_0$ for an arbitrary $\tilde{z}_i$ leaving only $N_f-1$
points to be determined by~\eqref{eqn29}.

As a consistency check of our solution we have also
considered the
decoupling of (infinitely) massive flavors and checked that our
formulas reduce to the case of pure $\nn=2$ theory without
flavors. We present some details of the computation in Appendix B.

We note that the solution satisfies a multiplication map. This map
was found in Ref.~\cite{Cachazo:2001jy,Cachazo:2002zk} for the case
without flavors and further generalized to the case with flavors in
Ref.~\cite{Balasubramanian:2003tv}. For any solution, with given
$N_c$, $N_1$, $\Delta k$, it follows from \eqref{eqn21},
\eqref{eqn22}, \eqref{eqn27} that we also have a solution for
$tN_c$, $t N_1$, $ t \Delta k$ with $t$ an integer, while at the
same time each $\tilde{z}_i$ is mapped onto $t$ copies of the same
$\tilde{z}_i$.

We finally remark that not all of the above discrete and continuous parameters
in the set $(N_1,\Dl k, k,\tau,z_0)$ give rise to different solutions. E.g. the
periodicity in $k$ is manifest and hence we have \eq (N_1,\Dl k,k)
\equiv (N_1,\Dl k, k+2N_c-N_f) \ . \eqx Also the periodicities of $N_1$
and $\Delta k$ can be found \eq\label{eqn:periodlk} (N_1,\Dl
k-(N_f-2N_c),k) \equiv (N_1,\Dl k, k-2N_c-2N_1) \ , \eqx
\eq\label{eqn:perion1} (N_1+N_f-2N_c,\Dl k,k) \equiv (N_1,\Dl k,
k-2\Dl k)\ . \eqx Equation~\eqref{eqn:periodlk} is easily seen noting
that $\infty_-$ from Eq.~\eqref{eqn22} changes by one. However, this
do not change the theta functions in the formula for $x$. Thus the
relation follows directly from Eq.~\eqref{eqn27}. On the other hand,
the periodicity in $N_1$ changes $\infty_-$ by $\tau$ and this means
that $x$ and the scale $B$ changes non-trivially and the relation
requires a calculation to check.

Similarly, we also expect modular transformations that change $\tau$
to $\tau+1$ or $-1/\tau$. This structure of the $\nn=1$ vacua will
be the subject of future investigation.

\section{Conclusions}

In this paper we have constructed an explicit solution of the
factorization problem of SW-curves for ${\cal{N}}=2$ supersymmetric
$U(N_c)$ gauge theory with $N_f$ fundamental flavors, when the gauge
symmetry is broken according to $U(N_c) \rightarrow U(N_1) \times
U(N_2)$. As a by-product we have rederived in a simpler way the
genus zero case of complete factorization first obtained in
Ref.~\cite{Demasure:2002jb}. Furthermore, for $N_f =0$ we get a
closed formula for
 the genus one case which was first solved in Ref.~\cite{Janik:2003hk}. We have also proven a
 theorem that holds for the general factorization. Finally, we have seen that our
 results can be applied to examine for what different sets of parameters one obtains the same
 factorized SW-curve, and hence the same physics. This is of relevance for the structure
 of  ${\cal{N}}=1$ vacua.

There are a number of interesting open issues that would be worth
studying. First of all, one could generalize the construction,
including the one without flavors, to other classical gauge groups.
Furthermore, it is still an open problem, also in the case with no
flavors present, to find a similar explicit solution for the case of
higher genus.  Another direction would be to consider the more
complicated case of quiver gauge theories where nonhyperelliptic
curves appear \cite{Casero:2003gr}. It would also be interesting to
apply the results here more directly to the Dijkgraaf-Vafa proposal
as was done for the exact results in the one-cut case
in Ref.~\cite{Ferrari:2002jp}.

Finally, we note that the existence of an exact solution is most interesting from
the point of view of studying in detail the global structure of ${\cal{N}}=1$ vacua
following Ref.~\cite{Cachazo:2002zk,Cachazo:2003yc}. The features
that were found there, including connected components of vacua and
possible dual descriptions of the same physics, are intimately
related to the discrete identifications between the parameters that
label the factorization solution. In this connection, it would be
interesting to see if our results can be used to further examine the
phase structure using the recent work of Ref.~\cite{Ashok:2006br}.
Here a fascinating connection was found between factorization of SW-curves
and Grothendieck's ``dessins d'enfants''. More generally, a
relation was conjectured between the programme of classifying these
dessins into Galois orbits and the problem of classifying special
phases of ${\cal{N}}=1$ vacua.

\bigskip

\noindent{}{\bf Acknowledgments.} We would like to thank Min-xin
Huang for discussion.

RJ was supported in part by Polish Ministry of Science and
Information Society Technologies grants 2P03B08225 (2003-2006),
1P03B02427 (2004-2007) and 1P03B04029 (2005-2008). Work partially
supported by the European Community's Human Potential Programme
under contracts MRTN-CT-2004-005616, `European Network on Random
Geometry' and MRTN-CT-2004-005104, `Constituents, fundamental forces
and symmetries of the universe'.

\appendix

\section{Factorization and Existence of the Meromorphic
One-Form}\label{app:theorem}

In this appendix we will prove that \emph{the SW-curve
with fundamental matter factorizes as in~\eqref{eqn8} if and only if
there exist a meromorphic one-form with only simple poles on a
hyperelliptic curve, $y_{\rm red}^2=F_{2(N_c-n)}(x)$, which has residue
$-N_c$ at infinity on the upper sheet, residue $N_c-N_f$ at infinity
on the lower sheet, residue $1$ at $-m_i$,
fulfills~\eqref{eqn12.5}, and, finally, has integral $\al$- and
$\be$-periods.}\footnote{That a meromorphic one-form with the given
poles, residues, and integral $\al$-periods exists is, of course,
always true.} Note that $N_c$, $N_f$, $m_i$ and $\La$ are
thought of as given.

This was proven in the case without fundamental matter
in \cite{Janik:2003hk}. The ideas here are much the same. The
proof is independent of the genus and is thus not confined to the
genus one curves considered above.

Let us first, for completeness, consider the easy part of the proof
and show that factorization of the SW-curve implies the
existence of the meromorphic 1-form on the reduced curve with the
prescribed properties.

\subsubsection*{Factorization implies integral 1-form}

In the first part of the proof we consider the factorized
SW-curve \eqref{eqn8} as given. Let us define
\begin{equation}\label{appeqn1}
    \om\equiv\bigg(\frac{P'_{N_c}(x)}{y(x)}+\frac{B'(x)}{2B(x)}-\frac{P_{N_c}(x)B'(x)}{2y(x)B(x)}\bigg)
    \mathrm{d}x \ .
\end{equation}
This is nicely a meromorphic one-form on the SW-curve~\eqref{eqn6}:
\begin{equation}\label{appeqn2}
    y^2=P_{N_c}(x)^2-4\La^{2N_c-N_f}B(x)\ ,
\end{equation}
where $B(x)=\prod_{i=1}^{N_f}(x+m_i)$. In fact,
using~\eqref{appeqn2} we get~\eqref{eqn12}:
\begin{equation}\label{appeqn3}
    \om=\mathrm{d}\log(P_{N_c}(x)+y(x)) \ ,
\end{equation}
which tells us that $\om$ has integral periods.
{} From~\eqref{appeqn1} we can also see that $\om$ has the right
poles and residues. However, since the curve is now
factorized according to ~\eqref{eqn8}:
\begin{equation}\label{appeqn4}
y^2=P_{N_c}(x)^2-4\La^{2N_c-N_f}B(x)=F_{2(N_c-n)}(x)H_{n}(x)^2=y_{\rm red}^2H_{n}(x)^2 \ ,
\end{equation}
we should check that we do not have poles at the zeroes of
$H_{n}$. Therefore let $x_0$ be a root in $H_n$. Then
by~\eqref{appeqn4} $x_0$ is a double root in $y^2$ and hence a
root in both $y^2$ and $d y^2/dx$. This gives
\begin{gather}\label{appeqn5}
     P_{N_c}(x_0)^2-4\La^{2N_c-N_f}B(x_0)=0 \ , \\
    2P_{N_c}(x_0)P'_{N_c}(x_0)-4\La^{2N_c-N_f}B'(x_0)=0 \ . \label{appeqn5.5}
\end{gather}
We will assume $B(x_0)\neq0$ and hence $P_{N_c}(x_0)\neq0$. Thus
we get from~\eqref{appeqn5} and~\eqref{appeqn5.5}
\begin{equation}\label{appeqn6}
    P'_{N_c}(x_0)-\frac{1}{2}P_{N_c}(x_0)\frac{B'(x_0)}{B(x_0)}=0  \ .
\end{equation}
Thus rewriting $\om$ from~\eqref{appeqn1} as
\begin{equation}\label{appeqn7}
    \om=\bigg(\frac{P'_{N_c}(x)-\frac{1}{2}P_{N_c}(x)\frac{B'(x)}{B(x)}}{H_n(x)}
    \frac{1}{y_{\rm red}(x)}+\frac{B'(x)}{2B(x)}\bigg)\mathrm{d}x \ ,
\end{equation}
we see by~\eqref{appeqn6} that the zeroes of $H_{n}$ are cancelled
and we do not get any poles from $H_n$. Thus we have proven that we
have a meromorphic one-form on the reduced curve, $y_{\rm red}$, with
the right poles and residues and with integral periods.\footnote{By
uniqueness (given the $\al$-periods) this must be $T(x)\mathrm{d}x$
from~\eqref{eqn11}.} That $\om$ fulfills~\eqref{eqn12.5} follows
directly from~\eqref{appeqn3} given that $P_{N_c}$ is normalized.

Before going to the second part of the proof let us get a little
inspiration from this case where we assume that the SW-curve factorizes. In the following if $x$ is a point in the upper
sheet then (with obvious abuse of notation) $\tilde{x}$ is the
corresponding point on the lower sheet. By~\eqref{appeqn1} we then
get (since $y(\tilde{x})=-y(x)$):
\begin{equation}\label{appeqn8}
    \om(x)+\om(\tilde{x})=\frac{B'}{B} \ .
\end{equation}
Now, let $a$ denote a branch point of $y_{\rm red}$. Then
integrating~\eqref{appeqn3} gives
\begin{gather}\label{appeqn9}
    P_{N_c}(a)e^{\int_a^x\om}=P_{N_c}(x)+y(x) \ , \\
    P_{N_c}(a)e^{\int_a^{\tilde{x}}\om}=P_{N_c}(x)-y(x) \ .\label{appeqn9.1}
\end{gather}
Performing the integrations entirely on the upper/lower sheet we
get from~\eqref{appeqn8}:
\begin{equation}\label{appeqn10}
    \int_a^{\tilde{x}}\om=-\int_a^x\om+\log\frac{B(x)}{B(a)}\ .
\end{equation}
Using this and choosing\footnote{There is really no sign choice in
$P_{N_c}$ since the coefficient of $x^{N_c}$ should be 1.}
$P_{N_c}(a)=2\La^{\frac{2N_c-N_f}{2}}\sqrt{B(a)}$ we find by
addition of~\eqref{appeqn9} and \eqref{appeqn9.1} that
\begin{equation}\label{appeqn11}
    P_{N_c}(x)=2\La^{\frac{2N_c-N_f}{2}}\sqrt{B(a)}
    \left(\frac{1}{2}e^{\int_a^x\om}+\frac{1}{2}\frac{B(x)}{B(a)}e^{-\int_a^x\om}\right) \ .
\end{equation}
This is independent of the chosen path of integration since $\om$
has integral periods.~\eqref{appeqn11} is the generalization for the
formula for the case without fundamental matter found
in \cite{Janik:2003hk}. There the generalization of the Chebyshev
polynomials from the one-cut case was found to be
$P(x)\propto\cosh(\int_a^x\om)$ which we get from~\eqref{appeqn11}
by taking $B(x)$ to be constant.

Let us now use the above considerations to complete the proof of the
theorem.

\subsubsection*{Integral 1-form on the reduced curve implies factorization}

For the second part of the proof we take as given a hyperelliptic
curve $y_{\rm red}$ and a meromorphic one-form $\om$ on $y_{\rm red}$ with
the prescribed poles, residues, integral periods, and
fulfilling~\eqref{eqn12.5}. In this case we simply define ($a$ is
again a branch point and we will, as above, assume that $B$ and
$y_{\rm red}^2$ do not share roots)
\begin{equation}\label{appeqn12}
    P_{N_c}(x)\equiv
    2\La^{\frac{2N_c-N_f}{2}}\sqrt{B(a)}\left(\frac{1}{2}e^{\int_a^x\om}+
    \frac{1}{2}\frac{B(x)}{B(a)}e^{-\int_a^x\om}\right) \ ,
\end{equation}
where we of course do not know if this is a polynomial. However, we
do know $P_{N_c}$ is well-defined in the sense that it is
independent of the choice of integration since $\om$ has integral
periods. To show that $P_{N_c}$ is indeed polynomial we will first
have to see that $\om$ fulfills~\eqref{appeqn8}. We know that we can
express $\om$ in the unique meromorphic one-forms $\om_{PQ}$ with
simple poles in $P$ and $Q$ with residues $+1$ and $-1$,
respectively, and zero $\al$-periods. In this way we can write
\begin{equation}\label{appeqn13}
    \om=-N_c\om_{\infty_+\infty_-}+\sum_i\om_{\widetilde{-m_i}
    \infty_-}+\textrm{holo. one-forms} \ ,
\end{equation}
as in~\eqref{eqn16} but where we now have a general genus and the
points are on the two sheets, not on the (generalized) torus. We
can now use that
\begin{eqnarray}\label{appeqn14}
    \om_{\infty_+\infty_-}(x)&=&-\frac{x^g}{y_{\rm red}(x)}\mathrm{d}x+\textrm{holo.
    one-forms}\ ,\\ \label{appeqn15}
    \om_{\tilde{P}\infty_-}(x)&=&\frac{1}{2}\frac{1}{x-P}\mathrm{d}x-\frac{1}{2}\frac{1}{x-P}\frac{x^{g+1}}{y_{\rm red}(x)}\frac{y_{\rm red}(P)}{P^{g+1}}\mathrm{d}x\\
    &&-\frac{1}{2}\left(1-\frac{y(P)}{P^{g+1}}\right)\frac{x^g}{y_{\rm red}(x)}
    \mathrm{d}x+\textrm{holo. one-forms}\ .\nonumber
\end{eqnarray}
Here $g$ is the genus of the reduced curve, i.e. if
$y_{\rm red}^2=F_{2(N_c-n)}(x)$ then $g=N_c-n-1$. Further a basis for
the holomorphic one-forms takes the form
\begin{equation}\label{appeqn16}
    \frac{x^i}{y_{\rm red}(x)}\mathrm{d}x,\quad i=0,\ldots,g-1 \ .
\end{equation}
Thus we can write $\om$ in~\eqref{appeqn13} as
\begin{equation}\label{appeqn16.5}
    \om(x)=\frac{1}{y_{\rm red}(x)}\left(R_g(x)-\frac{1}{2}\sum_i\frac{x^{g+1}}{x+m_i}
    \frac{y_{\rm red}(-m_i)}{(-m_i)^{g+1}}\right)\mathrm{d}x+\frac{1}{2}\sum_i\frac{1}{x+m_i}
    \mathrm{d}x \ ,
\end{equation}
where $R_g(x)$ is some polynomial of degree $g$ and we recognize
the last term as the expression $\tfrac{1}{2}B'(x)/B(x)\mathrm{d}x$. From
this~\eqref{appeqn8} is immediate. \eqref{appeqn8} then tells us
that $P_{N_c}$ is continuous across the cuts (it is of course by
definition continuous \emph{through} the cuts) since
using~\eqref{appeqn10}:
\begin{eqnarray}\nonumber
    P_{N_c}(\tilde{x})&=&2\La^{\frac{2N_c-N_f}{2}}\sqrt{B(a)}\left(\frac{1}{2}e^{-\int_a^x\om+\log\frac{B(x)}{B(a)}}+\frac{1}{2}\frac{B(x)}{B(a)}e^{\int_a^x\om-\log\frac{B(x)}{B(a)}}\right)\\
    &=&P_{N_c}(x) \ .\label{appeqn17}
\end{eqnarray}
This means that $P_{N_c}$ can be continued to a holomorphic
function in the (non-compact) complex plane with the possible
exception of the poles of $\om$ i.e. $\widetilde{-m_i}$. However,
the value of $P_{N_c}$ here is the same value as in $-m_i$
by~\eqref{appeqn17} and the are no poles in $-m_i$ at the upper
sheet. Thus we only have to care about the behavior of $P_{N_c}$
at infinity. Since $\int_a^x\om\sim N_c\log x$ for $x$ going to
infinity we get
\begin{equation}\label{appeqn18}
    \log P_{N_c}(x)\sim \log\big(e^{N_c\log x}+\frac{x^{N_f}}{B(a)}e^{-N_c\log
    x}\big)=\log\big(x^{N_c}+\frac{x^{N_f-N_c}}{B(a)}\big)\sim
    N_c\log x \ ,
\end{equation}
since $N_f\leq2N_c$. We can thus conclude that $P_{N_c}(x)$ is a
polynomial of degree $N_c$ as wanted. That $P_{N_c}$ is correctly
normalized follows by redoing the calculation in~\eqref{appeqn18}
also including the $x^0$-order and this time using the
assumption~\eqref{eqn12.5} and the derived
equation~\eqref{appeqn10}.\footnote{It is unclear if one really
has to assume~\eqref{eqn12.5}. In the case without fundamental
matter we simply rescale $x$ to get the correct value of $\La$.
However, in this case the rescaling also affects the masses thus
giving poles in the wrong places.}

Having established that $P_{N_c}$ is a polynomial it follows that
\begin{equation}\label{appeqn19}
    y^2\equiv P_{N_c}(x)^2-4\La^{2N_c-N_f}B(x) \ ,
\end{equation}
must also be polynomial. Now, all we need to prove is that
$y^2=y_{\rm red}^2H_{n}(x)^2$ for some polynomial $H_n$. Using
equation~\eqref{appeqn12} we get
\begin{equation}\label{appeqn20}
    y^2=4\La^{2N_c-N_f}B(a)\left(\frac{1}{4}e^{2\int_a^x\om}+\frac{1}{4}
    \frac{B(x)^2}{B(a)^2}e^{-2\int_a^x\om}-\frac{1}{2}\frac{B(x)}{B(a)}\right) \ .
\end{equation}
To see that $y^2$ contains $y_{\rm red}^2$ as a factor we first realize
that $a$, which was a root in $y_{\rm red}$, is also trivially a root in
$y^2$ by inserting $a$ in~\eqref{appeqn20}. Let now $b$ be any other
root in $y_{\rm red}^2$. We want to know the value of
$\exp(\pm2\int_a^b\om)$. To find these values we first note that the
$\al$- and $\be$-cycles on $y_{\rm red}$ can all be seen as curves from
one branch point to another on the upper sheet and then back again
(i.e. in the reverse direction) on the lower sheet (think of
continuous deformations of the curves in Figure~\ref{fig1}). The
curve in the integral $\int_a^b\om$ can then be seen as being put
together of the upper sheet parts of the $\al$- and $\be$-curves. We
can then write (using~\eqref{appeqn10} and explicitly writing
whether the integral is taken on the upper or the lower sheet):
\begin{eqnarray}
    \int_a^b\om\Big\lvert_{upper}&=&\frac{1}{2}\int_a^b\om\Big\lvert_{upper}-\frac{1}{2}\int_a^b\om\Big\lvert_{lower}+\frac{1}{2}\log\frac{B(b)}{B(a)}\nonumber\\
    &=&\frac{1}{2}\int_{\sum_{i}c_i\al_i+c'_i\be_i}\om+\frac{1}{2}\log\frac{B(b)}{B(a)}\ ,\label{appeqn21}
\end{eqnarray}
where the first integral in the last line simply is the half of a sum
of $\al$- and $\be$-periods of $\om$ ($c_i$ and $c'_i$ are $\pm1$
or zero). However, since we know that the periods are integral
(this is the crucial dependence on the integrality of the periods)
the first integral in the last line simply is an integer times $i
\pi$. Thus inserting in~\eqref{appeqn20} gives
\begin{equation}\label{appeqn22}
    y^2(b)=4\La^{2N_c-N_f}B(a)\left(\frac{1}{4}\frac{B(b)}{B(a)}+
    \frac{1}{4}\frac{B(b)}{B(a)}-\frac{1}{2}\frac{B(b)}{B(a)}\right)=0 \ .
\end{equation}
Thus the integrality of $\om$ gives us that $y^2$ contains
$y_{\rm red}^2$ as a factor, so that
\begin{equation}\label{appeqn23}
    y^2(x)=y_{\rm red}^2(x) Q(x) \ .
\end{equation}

To complete the proof we simply have to show that $Q(x)$ is the
square of a polynomial. First we note that
\begin{equation}\label{appeqn24}
    y(x)=2\La^{\frac{2N_c-N_f}{2}}\sqrt{B(a)}\left(\frac{1}{2}
    e^{\int_a^x\om}-\frac{1}{2}\frac{B(x)}{B(a)}e^{-\int_a^x\om}\right) \ ,
\end{equation}
since this nicely fulfills~\eqref{appeqn20} (we ignore the sign
choice). We can then calculate
\begin{multline}\label{appeqn25}
    \frac{1}{\sqrt{Q}}\frac{dy^2}{dx}=\frac{y_{\rm red}}{y}\frac{dy^2}{dx}=\frac{y_{\rm red}}{y}\left(2P_{N_c}P'_{N_c}-4\La^{2N_c-N_f}B'\right)\\
    \stackrel{\eqref{appeqn12},\eqref{appeqn24}}{=}\frac{y_{\rm red}}{y}\left(2P_{N_c}\left(y\frac{\om}{\mathrm{d}x}+2\La^{\frac{2N_c-N_f}{2}}\sqrt{B(a)}\frac{1}{2}\frac{B'}{B(a)}e^{-\int_a^x\om}\right)-4\La^{2N_c-N_f}B'\right)\\
    \stackrel{\eqref{appeqn12},\eqref{appeqn24}}{=}\frac{y_{\rm red}}{y}\left(2P_{N_c}\left(y\frac{\om}{\mathrm{d}x}+\frac{1}{2}\frac{B'}{B}(P-y)\right)-4\La^{2N_c-N_f}B'\right)\\
    \stackrel{\eqref{appeqn16.5},\eqref{appeqn19}}{=}\frac{y_{\rm red}}{y}\left(2P_{N_c}y\frac{1}{y_{\rm red}}\left(R_g-\frac{1}{2}\sum_i\frac{x^{g+1}}{x+m_i}\frac{y_{\rm red}(-m_i)}{(-m_i)^{g+1}}\right)+y^2\frac{B'}{B}\right)\\
    \stackrel{\eqref{appeqn23}}{=}\tilde{R}+y^2_{red}\sqrt{Q}\frac{B'}{B} \ ,
\end{multline}
where $\tilde{R}$ is some rational function. Solving this for $\sqrt{Q}$ gives
\begin{equation}\label{appeqn26}
    \sqrt{Q(x)}=\frac{1}{\tilde{R}(x)}\left(\frac{dy^2}{dx}-y^2_{red}\frac{B'}{B}Q\right) \ .
\end{equation}
Since $\sqrt{Q}$ is the square-root of a polynomial and the right
hand side is a rational function we can finally conclude that
$\sqrt{Q}$ is a polynomial, thus finishing the proof.

The given one-form $\om$ must then -- by uniqueness given the
$\al$-periods -- be $T(x)\mathrm{d}x$ from equation~\eqref{eqn11}
and thus~\eqref{eqn11.1} applies.

\section{Flavor decoupling}

We can obtain the case without fundamental flavors by taking the
limits described after equation~\eqref{eqn7}. This means that we
should take $\La\rightarrow0$ and $m_i\rightarrow\infty$ for all
$i$ while keeping constant:
\begin{equation}\label{eqn31}
    \La^{2N_c-N_f}\prod_i m_i\equiv\Lanew^{2N_c} \ ,
\end{equation}
where $\Lanew$ is the new scale for the theory without flavors. The
limit means that for all $i$ we have $\tilde{z}_i\rightarrow
\infty_-$. However, $\infty_-$ is itself changed by changing the
$\tilde{z}_i$s so from~\eqref{eqn22} we get the consistency
equation
\begin{equation}\label{eqn32}
    \infty_-=\frac{(N_1-N_c)\ta-\De k-N_c+N_f\infty_-}{N_f-2N_c} \ ,
\end{equation}
which, as could be expected, is solved as
\begin{equation}\label{eqn33}
    \infty_-=\frac{(N_1-N_c)\ta-\De k-N_c}{-2N_c} \ .
\end{equation}
For $x(z)$ in~\eqref{e.xz} all we should do
then is to find the formula for $\scaleb$ after the limit has been
taken. Using~\eqref{e.xz} in $m_i=-x(\tilde{z}_i)$ we can rewrite
equation~\eqref{eqn31} as
\begin{multline}\label{eqn34}
    \lim\frac{\La^{2N_c-N_f}}{\prod_i\tha(\tilde{z}_i-\infty_-+\tfrac{1+\ta}{2})}\bigg(-\scaleb\frac{\tha(\infty_--z_0+\tfrac{1+\ta}{2})\tha(\infty_--1-\ta+z_0+\tfrac{1+\ta}{2})}{\tha(\infty_--\infty_++\tfrac{1+\ta}{2})}\bigg)^{N_f}\\
    =\Lanew^{2N_c} \ .
\end{multline}
This means we can solve for $\lim
\La^{2N_c-N_f}/\prod_i\tha(\tilde{z}_i-\infty_-+\tfrac{1+\ta}{2})$
and the result can be used in~\eqref{eqn27} to obtain:
\begin{multline}\label{eqn35}
    \scaleb=e^{\frac{i2\pi
    k}{2N_c-N_f}}\frac{\tha(\infty_+-\infty_-+\tfrac{1+\ta}{2})^2}{\tha(\infty_+-z_0+\tfrac{1+\ta}{2})\tha(\infty_+-1-\ta+z_0+\tfrac{1+\ta}{2})}\\
    \times \tha(\infty_+-\infty_-+\tfrac{1+\ta}{2})^{\frac{N_f}{2N_c-N_f}}e^{2\pi i(\infty_+-\infty_-)\frac{N_c+N_1}{2N_c-N_f}}e^{-\pi
    i\frac{N_f}{2N_c-N_f}}\\
    \times \Lanew^{\frac{2N_c}{2N_c-N_f}}
    \bigg(-\frac{\tha(\infty_--\infty_++\tfrac{1+\ta}{2})}{\scaleb\tha(\infty_--z_0+
    \tfrac{1+\ta}{2})\tha(\infty_--1-\ta+z_0+\tfrac{1+\ta}{2})}\bigg)^{\frac{N_f}{2N_c-N_f}} \ .
\end{multline}
We can then solve for $\scaleb$ to obtain
\begin{multline}\label{eqn36}
    \scaleb=\Lanew e^{\frac{i2\pi
    k}{2N_c}}\frac{\tha(\infty_+-\infty_-+\tfrac{1+\ta}{2})^2}{\tha(\infty_+-z_0+\tfrac{1+\ta}{2})\tha(\infty_+-1-\ta+z_0+\tfrac{1+\ta}{2})}\\
\times e^{2\pi i(\infty_+-\infty_-)\frac{N_c+N_1}{2N_c}} \ .
\end{multline}
This gives the solution together with~\eqref{e.xz}:
\begin{equation}\label{eqn37}
    x(z)=\scaleb\frac{\tha(z-z_0+\tfrac{1+\ta}{2})
    \tha(z-1-\ta+z_0+\tfrac{1+\ta}{2})}{\tha(z-\infty_++\tfrac{1+\ta}{2})\tha(z-\infty_-+\tfrac{1+\ta}{2})} \ ,
\end{equation}
and the limit of~\eqref{eqn20}:
\begin{equation}\label{eqn38}
    \om=N_c\mathrm{d}\log\frac{\tha(z-\infty_-+\tfrac{1+\ta}{2})}{\tha(z-\infty_++\tfrac{1+\ta}{2})}+2\pi i N_1\mathrm{d}z \ .
\end{equation}
This is exactly what we would get if we solved the factorization
problem without fundamental matter directly.



\providecommand{\href}[2]{#2}\begingroup\raggedright\endgroup

\end{document}